\documentclass[aps,pra,twocolumn,showpacs,groupedaddress]{revtex4}
\usepackage{graphicx}

\begin{document}


\title{Wave mixing of optical pulses and
 Bose-Einstein condensates}
\author{Han Pu, Weiping Zhang, and Pierre Meystre}
\affiliation{Optical Sciences Center, University of Arizona,
Tucson, AZ 85721}

\date{\today}

\begin{abstract}
We investigate theoretically the four-wave mixing of optical and
matter waves resulting from the scattering of a short light pulse
off an atomic Bose-Einstein condensate, as recently demonstrated
by D. Schneble {\em et al.} [ Science {\bf 300}, 475 (2003)]. We
show that atomic ``pair production'' from the condensate results
in the generation of both forward- and backward-propagating matter
waves. These waves are characterized by different phase-matching
conditions, resulting in different angular distributions and
temporal evolutions.
\end{abstract}

 \maketitle

The interaction between optical fields and atomic Bose-Einstein
condensates has attracted much recent attention, due of its
importance in the preparation, manipulation, and detection of
condensates, as well as because of its interest in fundamental
studies of the nonlinear interaction between Maxwell and
Schr{\"o}dinger waves. A number of phenomena have already been
studied both theoretically and experimentally, including
matter-wave superradiance \cite{super1,super2,super3}, coherent
matter-wave amplification \cite{amp1,amp2,amp3} and Bragg
spectroscopy \cite{bragg1,bragg2,bragg3}.

In a trailblazing experiment \cite{super1}, a cw laser beam shined
on a cigar-shaped condensate resulted in the generation of a
fan-like pattern of momentum sidemodes of the condensate. The
initiation of this pattern can be understood in terms of a
four-wave mixing process involving two optical fields -- the laser
field and a so-called end-fire mode; and two matter-wave modes --
the condensate and a mode of momentum such that energy-momentum
conservation (or phase matching) is satisfied. We recall here that
the end-fire mode, first predicted by Dicke \cite{Dicke} in his study of
superradiance, corresponds to the privileged direction for
spontaneous emission along the long axis of the condensate.
Because of momentum conservation, it is clear that the generated
sidemodes must be in the forward direction at a 45$^\circ$ angle
between the direction of the incident laser and the long axis of
the condensate. The subsequent generation of further sidemodes
results simply from wave mixing involving an already excited
sidemode instead of the initial condensate at rest.

Recently, that same MIT group \cite{ketterle} reported the results
of an experiment where the incident cw laser was replaced by an
optical pulse. This led to the remarkable result that for short
enough pulses, backward-scattered atoms (with a momentum component
antiparallel to the direction of the pump field) were also
observed. Moreover, the backward peaks exhibited a slightly
different angular distribution compared with the forward peaks.
The qualitative difference in diffraction patterns for the short-
and long-pulse regimes was attributed to the transition from the
Raman-Nath to the Bragg regime of diffraction, that is, to the
onset of energy-momentum conservation for long enough interaction
times.

In this Letter, we present a theoretical description of this
experiment based on an extension of the quasi-mode approach of
Ref. \cite{super2}. We give a full dynamical treatment of both the
vacuum photon modes and the condensate sidemodes that interprets
the diffraction pattern in terms of atom-photon wave mixing and
show explicitly that the counter-propagating (backward and
forward) matter-wave quasi-modes result from the
quantum-correlated parametric excitation of atomic pairs. The
difference in their angular distribution and dynamics results from
their distinct phase-matching conditions.

A general theoretical framework to describe the interaction of
ultracold atoms with light waves was presented by Zhang and Walls
in Ref. \cite{zhang}. In the situation at hand, the atomic system,
assumed to be at zero temperature, interacts with a far
off-resonant classical laser field of (real) Rabi frequency
$\Omega_L$, wave vector ${\bf k}_L$ and frequency $\omega_L=ck_L$,
as well as with a continuum of electromagnetic field modes of wave
vector ${\bf k}$ and polarization $\lambda$ characterized by the
bosonic annihilation operators $B_{{\bf k}\lambda}$. In the case
of large atom-laser detunings $\Delta$, we are justified in
adiabatically eliminating the excited atomic levels, leaving us
with just the bosonic ground-state matter-wave field operator
$\psi_1({\bf r},t)$. We furthermore perform the rotating wave
approximation and express the dynamics of the coupled
atoms-radiation system in a frame rotating at the pump laser
frequency to find
\begin{eqnarray}
i\hbar \frac{\partial \psi_1}{\partial t} &=& H_0({\bf r}) \psi_1
+ \frac{\hbar \Omega_L}{2\Delta} \sum_{{\bf k}, \lambda}
\left[ g^*_{{\bf k}\lambda} B_{{\bf k}\lambda}^{\dagger} \right.\nonumber \\
&\times& \left. e^{-i({\bf k}-{\bf k}_L)
\cdot {\bf r} -i\omega_L t}+h.c. \right] \psi_1 ,\label{12} \\
i\hbar \frac{\partial B_{{\bf k} \lambda}}{\partial t} &=& \hbar
\omega_k  B_{{\bf k} \lambda}+\frac{\hbar \Omega_L}{2\Delta}
g^*_{{\bf k} \lambda} e^{-i \omega_L t} \nonumber \\
&\times& \int d^3r\, e^{-i ({\bf k}-{\bf k}_L) \cdot {\bf r}}
\psi_1^{\dagger}({\bf r},t) \psi_1({\bf r}, t), \label{B2}
\end{eqnarray}
where $g_{{\bf k} \lambda}=i\sqrt{2\pi \omega_k/(\hbar V)} {\bf d}
\cdot {\bf e}_{{\bf k}\lambda}$ is the coupling strength of the
atom with the corresponding vacuum mode ($V$ is the quantization
volume, ${\bf e}_{{\bf k}\lambda}$ the photon polarization unit
vector and ${\bf d}$ the atomic dipole moment). In Eq.~(1), the
atomic Hamiltonian  $H_0({\bf r})$ includes the usual kinetic and
trapping potential terms, the ac Stark shift arising from the pump
laser as well as nonlinear atom-atom interactions resulting e.g.
from two-body collisions.

In the absence of electromagnetic fields, the condensate at zero
temperature is taken to be in the ground state $\varphi_0({\bf
r})$ of $H_0$, with $(H_0 - \hbar \mu)\varphi_0({\bf r}) =0$,
$\mu$ being the chemical potential. The interaction of this
condensate with light shifts the atomic momentum by the photon
recoil, and transfers condensate atoms to states of the form
$\varphi_0({\bf r}) \exp(i {\bf q}\cdot {\bf r})$. For condensates
large compared to an optical wavelength, these states are still
approximate eigenstates of $H_0$, with eigenenergy shifted by the
recoil frequency $\omega_q = \hbar q^2/2M$, i.e., $
H_0\,[\varphi_0({\bf r}) \exp(i{\bf q} \cdot {\bf r})] \approx
\hbar (\mu + \omega_q) \,\varphi_0({\bf r}) \exp(i{\bf q} \cdot
{\bf r})$, an approximation similar to the slowly varying envelope
approximation in optics. Following Ref. \cite{super2}, this
suggests expanding $\psi_1({\bf r},t)$ as
\begin{equation}
\psi_1 ({\bf r},t)= e^{-i\mu t}\,\sum_{{\bf q}} \varphi_0({\bf r})
e^{i({\bf q} \cdot {\bf r}- \omega_q t)} c_{\bf q},\label{exp}
\end{equation}
where the atomic quasi-mode field operators approximately obey
bosonic commutation relations, $[c_{\bf q}, c^\dagger_{{\bf q}'}]
= \delta_{{\bf q},{\bf q}'}$. Inserting this expansion into
Eq.~(\ref{12}) gives
\begin{widetext}
\begin{equation}
i\hbar \dot{c}_{\bf q} =\frac{\hbar \Omega_L}{2\Delta} \sum_{{\bf
k},\lambda} \sum_{{\bf q}'} \left[ g_{{\bf k}\lambda}^*
B^{\dagger}_{{\bf k}\lambda} e^{-i\omega_L t} \Pi_{{\bf q}-{\bf
q}'}({\bf k}) + g_{{\bf k}\lambda} B_{{\bf k}\lambda} e^{i
\omega_L t} \Pi^*_{{\bf q}'-{\bf q}}({\bf k}) \right]
e^{i(\omega_q-\omega_{q'}) t}\,c_{{\bf q}'} ,\label{cq}
\end{equation}
where $\Pi_{\bf q}({\bf k}) = \int d^3r \, |\varphi({\bf r})|^2
\exp[-i({\bf q}+{\bf k}-{\bf k}_L) \cdot {\bf r}]$ is the spatial
Fourier transform of the condensate density profile. Consistently
with the slowly-varying approximation implicit in the expansion
(\ref{exp}), it satisfies the approximate relation $\Pi_{\bf
q}^*({\bf k}) \Pi_{{\bf q}'}({\bf k}) \approx |\Pi_{\bf q}({\bf
k})|^2\, \delta_{{\bf q},{\bf q}'}$, a condition that indicates
that the discretization of ${\bf q}$ is such that there is
neglibible overlap between shifted momentum wave functions. This
property is used repeatedly in the following.

We now proceed by formally integrating Eq.~(\ref{B2}) to find
\begin{eqnarray}
B_{{\bf k}\lambda}(t) &=& B_{{\bf k}\lambda}(0) e^{-i\omega_k t}
-i\frac{\Omega_L}{2\Delta} g^*_{{\bf k} \lambda} e^{-i\omega_L t}
\int_0^t d\tau\,e^{-i(\omega_k - \omega_L) \tau} \,\int d^3r \,
e^{-i({\bf k}-{\bf k}_L) \cdot {\bf r}} \psi_1^{\dagger}({\bf r},
t-\tau) \psi_1({\bf r}, t-\tau) \nonumber \\
&=& B_{{\bf k}\lambda}(0) e^{-i\omega_k t} -i\frac{\pi
\Omega_L}{\Delta} g^*_{{\bf k} \lambda} e^{-i\omega_L t}
\sum_{{\bf q}_1,{\bf q}_2} \Pi_{{\bf q}_1-{\bf q}_2}({\bf k})
\delta(\omega_k - \omega_L+\omega_{q_1}-\omega_{q_2})
e^{i(\omega_{q_1}-\omega_{q_2})t}\,c^{\dagger}_{{\bf q}_1}(t)
c_{{\bf q}_2}(t), \label{B3}
\end{eqnarray}
\end{widetext}
where we have used Eq.~(\ref{exp}) and have adopted the Markov
approximation \cite{notemarkov} by replacing $c^{\dagger}_{{\bf
q}_1}(t-\tau)$ and $ c_{{\bf q}_2}(t-\tau)$ with
$c^{\dagger}_{{\bf q}_1}(t)$ and $c_{{\bf q}_2}(t)$ respectively,
and setting the upper limit of integration over $\tau$ to infinity
\cite{note}.

It is clear by inspection of Eqs. (\ref {cq}) and (\ref{B3}) that
the system dynamics can be interpreted in terms of four-wave
mixing processes involving two matter-wave fields and two optical
fields. For example, the dynamics of the atomic quasi-mode ${\bf
q}$ results from the combined effects of the laser and a vacuum
field mode together with an additional atomic quasi-mode ${\bf
q}'$, while the dynamics of a scattered photon mode $\{{\bf k},
\lambda \}$ results in turn from the combined effects of the laser
field and two atomic modes.

For optical pulses of short duration, we can assume that the
condensate population is undepleted and replace the operators
$c_0$ and $c_0^{\dagger}$ by the $c$-number $\sqrt{N_0}$, where
$N_0$ is the number of condensate atoms. Inserting then
Eq.~(\ref{B3}) into Eq.~(\ref{cq}) results in,
\begin{eqnarray}
\dot{c}_{\bf q} =  {\cal A}_{\bf q}(t) c_{\bf q} + {\cal B}_{\bf
q}(t) c_{-{\bf q}}^{\dagger} + \Gamma_{\bf q}(t), \label{cq1}
\end{eqnarray}
where ${\bf q} \neq 0$ and we have only kept the linear terms
involving $c_{\bf q}$ and $c_{\bf q}^{\dagger}$, consistently with
the undepleted pump approximation. Here
\begin{eqnarray*}
{\cal A}_{\bf q}(t) &=& \frac{N_0 \pi \Omega_L^2}{2\Delta^2}
\sum_{{\bf k},\lambda} |g_{{\bf k},\lambda}|^2 \left [ |\Pi_{\bf
q}({\bf k})|^2 \delta(\omega_k-\omega_L+\omega_q) \right . \\
&-& \left .|\Pi_{-{\bf q}}({\bf k})|^2 \delta
(\omega_k-\omega_L-\omega_q) \right ],
\end{eqnarray*}
and
$
{\cal B}_{\bf q}(t) = -{\cal A}_{-{\bf q}} \exp(2i\omega_q t).
$

The Langevin noise operators $\Gamma_{\bf q}(t)$, which account
for the vacuum electromagnetic fluctuations through the initial
photon operators $B_{{\bf k}\lambda}(0)$ and $B_{{\bf
k}\lambda}^{\dagger}(0)$ and are responsible for the initiation of
the scattering process \cite{super2}, have the form $\Gamma_{\bf
q}(t)= \exp(i\omega_q t) \,[f_{\bf q}^{\dagger}(t)-f_{-{\bf
q}}(t)]$, where
\[f_{\bf q}(t) =
i\frac{\sqrt{N_0} \Omega_L}{2\Delta} \sum_{{\bf k},\lambda}
g_{{\bf k}\lambda} B_{{\bf k}\lambda}(0)
e^{-i(\omega_k-\omega_L)t} \Pi_{\bf q}^*({\bf k}).\]

\begin{figure}
\includegraphics*[width=0.78\columnwidth,
height=0.66\columnwidth]{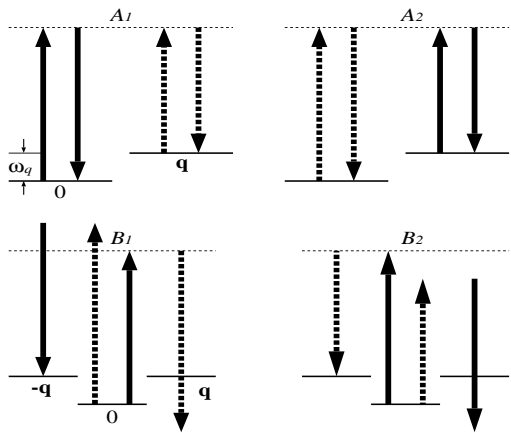} \caption{Plots $A_1$ and $A_2$
($B_1$ and $B_2$) are the two scattering processes represented by
the coefficient ${\cal A}_{\bf q}$ (${\cal B}_{\bf q}$). ``0'' and
``$\pm {\bf q}$'' label the atomic momentum states with ``0''
being the condensate mode. The solid and dashed arrows represent
the pump and scattered photon, respectively.} \label{fig1}
\end{figure}

The coefficients ${\cal A}_{\bf q}$ and ${\cal B}_{\bf q}$ each
contains two terms, corresponding to the four processes
illustrated in Fig.~\ref{fig1}. The two processes described by
${\cal A}_{\bf q}$ are resonant, while those described by ${\cal
B}_{\bf q}$ are characterized by an energy mismatch $2\hbar
\omega_q$, as evident from the expression for ${\cal B}_{\bf q}$.
Fig.~\ref{fig1} shows that ${\cal B}_{\bf q}$ describes a pair
production process where two condensate atoms are scattered into
the quasi-modes ${\bf q}$ and $-{\bf q}$. Such a process is known
in quantum optics to result in the production of entangled or
squeezed particle pairs. This process, which leads to the
generation of backward-scattered atoms, will clearly occur for
interaction times short enough that energy-momentum conservation
(or phase-matching) is not yet established. This explains why
back-scattering was not observed in the quasi-cw experiments of
Ref. \cite{super1}. Furthermore, ${\cal A}_{\bf q}(t)$ is positive
(negative) for forward- (backward-)scattered atoms. Hence, the
onset of backward scattering relies fully on the pair production
characterized by ${\cal B}_{\bf q}(t)$.

Equation~(\ref{cq1}) readily gives the quasi-mode populations
$n_{\bf q} = \langle c_{\bf q}^{\dagger} c_{\bf q} \rangle$ as
\begin{eqnarray}
\dot{n}_{\bf q} &=& 2{\cal A}_{\bf q}n_{\bf q} + {\cal B}_{\bf q}
m_{\bf q}^* + {\cal B}_{\bf q}^* m_{\bf q} +{\cal N}_{\bf q},
\label{nq} \\
\dot{m}_{\bf q}& =& ({\cal A}_{\bf q} + {\cal A}_{-{\bf q}})m_{\bf
q} + {\cal B}_{-{\bf q}} n_{\bf q} + {\cal B}_{\bf q} n_{-{\bf
q}}+{\cal M}_{\bf q}, \label{mq}
\end{eqnarray}
where the appearance of the anomalous density $m_{\bf q}=m_{-{\bf
q}}=\langle c_{-{\bf q}} c_{\bf q} \rangle$  is a clear signature
of the quantum correlations between the modes ${\bf q}$ and $-{\bf
q}$, and the quantities ${\cal N}_{\bf q}$ and ${\cal M}_{\bf q}$
result from the noise operators $\Gamma_{\bf q}$. Their explicit
forms can be obtained using the quantum regression theorem
\cite{regression} as
\begin{eqnarray*}
{\cal N}_{\bf q}=\frac{N_0 \Omega_L^2}{2\Delta^2} \sum_{{\bf
k},\lambda} |g_{{\bf k}\lambda}|^2 |\Pi_{\bf q}({\bf k})|^2
\,\delta(\omega_k-\omega_L+\omega_q) ,
\end{eqnarray*}
and ${\cal M}_{\bf q}=-({\cal N}_{\bf q} + {\cal N}_{-{\bf q}})\,
\exp(i2\omega_q t)/2 $.

For completeness, we also give the expression for the number of
scattered photons in mode $\{{\bf k}, \lambda \}$,
\begin{widetext}
\begin{equation}
\langle B_{{\bf k}\lambda}^{\dagger}(t) B_{{\bf k}\lambda}(t)
\rangle \approx \frac{N_0 \Omega_L^2 |g_{{\bf
k}\lambda}|^2}{4\Delta}  \sum_{{\bf q} \neq 0} \left\{ |\Pi_{\bf
q}({\bf k})|^2 \delta (\omega_k-\omega_L+\omega_q) \,[n_{\bf
q}(t)+1] +|\Pi_{-{\bf q}}({\bf k})|^2 \delta
(\omega_k-\omega_L-\omega_q) \,n_{\bf q}(t) \right\}.\label{bk}
\end{equation}
\end{widetext}

\begin{figure}
\includegraphics*[width=0.8\columnwidth,
height=0.45\columnwidth]{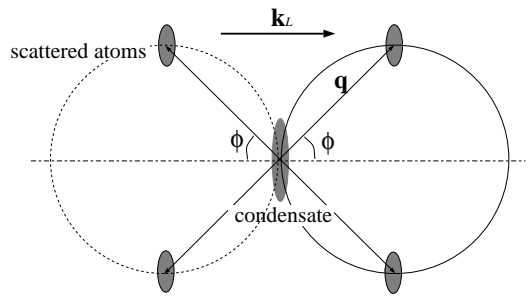} \caption{Schematic of the
dipole emission pattern. The circle with solid (dashed) line is
traced out by the momenta {\bf q} of the forward- (backward-)
scattered atoms.} \label{fig2}
\end{figure}

Equations~(\ref{nq}) and (\ref{mq}) form a closed set that can be
solved numerically. The atomic recoil frequencies $\omega_q$ are
many orders of magnitude smaller than the laser frequency
$\omega_L$, so that conservation of energy requires that the wave
numbers of the scattered photons satisfy $k \simeq k_L$.
Furthermore, for the large condensates considered here, the
slowly-varying amplitude approximation mentioned earlier implies
that $\Pi_{\bf q}({\bf k})$ is only significantly different from
zero for ${\bf k}+{\bf q}-{\bf k}_L \simeq 0$, indicating momentum
conservation for the scattering process. Under such conditions,
the atoms are scattered from the condensate in two dipole emission
halos, represented schematically by the two circles of radius
$k_L$ in Fig.~\ref{fig2}.

We evaluate the coefficients in Eqs.~(\ref{nq}) and (\ref{mq}), by
following the procedure of Ref. \cite{super2} to find ${\cal
A}_{\bf q} = {\cal B}_{\bf q} \exp(-i2\omega_q t)= \gamma
\,(\Omega_{\bf q} - \Omega_{-{\bf q}})$, ${\cal N}_{\bf q} =
2\gamma \Omega_{\bf q}$ and ${\cal M}_{\bf q} = -\gamma
\exp(i2\omega_q t) \,(\Omega_{\bf q} + \Omega_{-{\bf q}} )$, where
$\gamma=N_0 \Omega_L^2 k_L^3 |{\bf d}|^2/(4\pi \hbar \Delta^2)$
characterizes the strength of the atom-photon interaction, and
\[\Omega_{\bf q}=\frac{4\pi}{k_L^2 w^2}\left[ \cos^2 \theta_{\bf
k}+ (l/w)^2 \sin^2 \theta_{\bf k} \right]^{-1/2} .\] Here, $w$ and
$l$ are the dimensions of the cylindrically symmetric condensate
along the radial and axial directions, respectively, and
$\theta_{\bf k}$ is the angle between ${\bf k}$($={\bf k}_L-{\bf
q}$) and the long axis of the condensate. For a cigar-shaped
condensate, $l \gg w$, the scattered photons are predominantly
along that axis, Dicke's end-fire mode, so that most of the
scattered atoms gain momenta at roughly 45$^\circ$ from the long
axis of the condensate \cite{super1,super2,ketterle}.

\begin{figure}
\includegraphics*[width=0.8\columnwidth,
height=1.0\columnwidth]{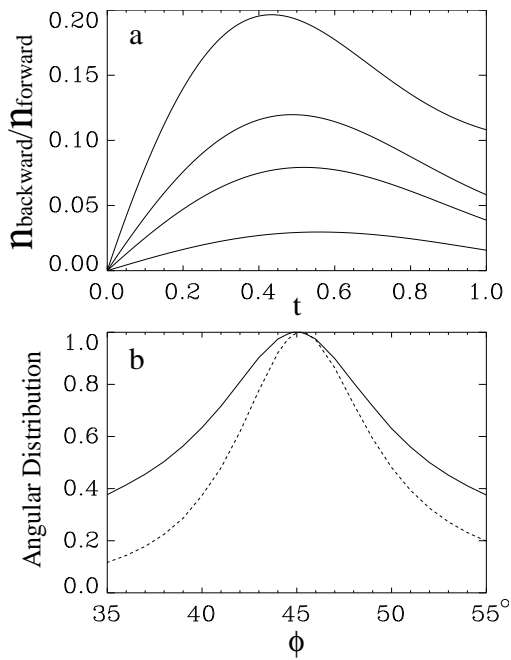} \caption{a) Ratio of the
backward- and forward-scattered atom numbers at angle
$\phi=45^\circ$ as a function of time, in units of $1/\omega_r$,
where $\omega_r=\hbar k_L^2/(2M)$. The lines from top to bottom
correspond to $\gamma=500$, 250, 150 and 50$\omega_r$. b) Angular
distribution in arbitrary units of the forward- (solid line) and
backward-scattered (dashed line) atoms. Here $\gamma=50\omega_r$
and $t=0.7/\omega_r$. The results are obtained by numerically
integrating Eqs.~(\ref{nq}) and (\ref{mq}) with initial conditions
$n_{\bf q}=m_{\bf q}=0$ using a 4$^{th}$-order Runge-Kutta
method.} \label{fig3}
\end{figure}

Figure~\ref{fig3}a shows the ratio of the backward- and
forward-scattered atom numbers along this 45$^\circ$ angle. This
ratio is always less than 1 since the backward scattering process
has an energy mismatch $\Delta E=2\hbar \omega_q$ while the
forward process is resonant \cite{ketterle}. Fig.~\ref{fig3}b
shows the angular distribution of the scattered atoms. The
forward-scattered atoms form an almost perfect symmetric
distribution about the 45$^\circ$ angle. This must be contrasted
to the backward-scattered atoms, whose distribution shows an
asymmetry, with more atoms scattered into $\phi> 45^\circ$ than
into $\phi < 45^\circ$ (see Fig.~\ref{fig2}): Backward scattering
favors larger angles because the energy mismatch is smaller for
$\phi> 45^\circ$ than that for $\phi< 45^{\circ}$, since $\Delta E
=2\hbar \omega_q \propto \cos^2 \phi$.  A similar asymmetry in the
behavior of the forward- and backward-scattered atoms was observed
experimentally \cite{ketterle}, and was attributed to the fact
that the two scattering processes occur mainly at different
locations along the condensate. Because the Markov approximation
neglects the effects of retardation in the description of the
end-fire mode dynamics \cite{ressayre}, that mechanism is absent
from our analysis. In practice, both phase matching and spatial
effects are probably at play, and they could in principle be
distinguished by varying the time of flight before detection
\cite{wolfgang}.

In summary, we have presented a theoretical investigation of a
zero-temperature condensate interacting with a far off-resonant
laser pulse. Our work is valid for interaction times short enough
that the condensate remains undepleted and one can retain only
linear terms involving atomic modes of finite momenta. Our
analysis shows that while the resonant Raman scattering
illustrated in the plots $A_{1,2}$ of Fig.~\ref{fig1} is the
dominant generation mechanism of forward-scattered atoms, the
backward-scattered atoms result from the quantum-correlated pair
production processes of plots $B_{1,2}$ in Fig.~\ref{fig1}
\cite{corr}. The lack of phase-matching for such pair production
processes results in distinct angular distributions for the
backward- and forward-scattered atomic peaks, in good qualitative
agreement with the experiment of Ref. \cite{ketterle}.

Schneble {\em et al.} interpret the existence of
backward-scattered atoms as resulting from Kapitza-Dirac
diffraction of matter waves off the optical gratings formed by the pump laser and the end-fire modes \cite{ketterle}. While this point-of-view is consistent with our correlated pair production picture, we emphasize that the strength of the optical
grating depends on the intensity of the end-fire modes, which in
turn depends on the population of the atomic quasi-modes [see
Eq.~(\ref{bk})]. It is this interplay of the optical and atomic
fields that renders the backward- and forward-scattered atoms
correlated. It also shows that the two diffraction processes of
the system---optical diffraction off the atomic grating and atomic diffraction off the optical grating---are coherently mixed and hence inseparable. For the same reason, the two
counter-propagating optical end-fire modes must also exhibit
quantum correlations.

We thank Prof. Ketterle for sending us a preprint of
Ref. \cite{ketterle} before publication and for discussions. This work is
supported in part by the US Office of Naval Research under
Contract No. N00014-03-1-0388, by the National Science Foundation under Grant No. PHY00-98129, by the US Army Research Office, by the NASA Microgravity Fundamental Physics Program and by the Joint Services Optics Program.


\begin{references}
\bibitem{notemarkov}The Markov approximation is justified here since Eq.~(\ref {B3}) describes
the evolution of the electromagnetic field, and photons escape the
condensate in a time short compared to all other times of
interest. It does however neglect the effects of retardation, a point to which we return later.
\bibitem{note}This is valid as long as the upper limit of
integration $t \gg 1/\omega_L$ and we note that $1/\omega_L$ is
much short than any time scale of condensate dynamics.
\bibitem{corr}Quantum correlations
between forward-scattered atoms were previously mentioned by
M\"{u}stecaplio\v{g}lu and You in Ref. \cite{super3}.
\end{references}
\end{document}